\documentclass[twocolumn]{aastex63}
\pdfoutput=1 

\usepackage{hyperref}
\hypersetup{colorlinks,allcolors=blue}

\shorttitle{A Clear Atmosphere for WASP-62$\lowercase{\text b}$}
\shortauthors{Alam et al.}

\begin{document}

\title{Evidence of a Clear Atmosphere for WASP-62$\lowercase{\text b}$: the Only Known Transiting \\ Gas Giant in the JWST Continuous Viewing Zone}

\correspondingauthor{Munazza K. Alam}
\email{munazza.alam@cfa.harvard.edu}

\author[0000-0003-4157-832X]{Munazza K. Alam}
\altaffiliation{National Science Foundation Graduate Research Fellow}
\affiliation{Center for Astrophysics $|$ Harvard {\rm \&} Smithsonian, 60 Garden Street, Cambridge, MA 02138, USA}

\author[0000-0003-3204-8183]{Mercedes L\'opez-Morales} 
\affil{Center for Astrophysics $|$ Harvard {\rm \&} Smithsonian, 60 Garden Street, Cambridge, MA 02138, USA}

\author[0000-0003-4816-3469]{Ryan J. MacDonald}
\affil{Department of Astronomy and Carl Sagan Institute, Cornell University, Ithaca, NY 14853, USA}

\author[0000-0002-6500-3574]{Nikolay Nikolov}
\affil{Space Telescope Science Institute, 3700 San Martin Drive, Baltimore, MD 21218, USA}

\author[0000-0002-4207-6615]{James Kirk}
\affil{Center for Astrophysics $|$ Harvard {\rm \&} Smithsonian, 60 Garden Street, Cambridge, MA 02138, USA}

\author[0000-0002-8515-7204]{Jayesh M. Goyal}
\affil{Department of Astronomy and Carl Sagan Institute, Cornell University, Ithaca, NY 14853, USA}

\author[0000-0001-6050-7645]{David K. Sing}
\affil{Department of Earth {\rm \&} Planetary Sciences, Johns Hopkins University, Baltimore, MD, USA} 
\affil{Department of Physics {\rm \&} Astronomy, Johns Hopkins University, Baltimore, MD, USA}

\author[0000-0003-4328-3867]{Hannah R. Wakeford}
\affil{School of Physics, University of Bristol, HH Wills Physics Laboratory, Tyndall Avenue, Bristol BS8 1TL, UK}

\author[0000-0002-4227-4953]{Alexander D. Rathcke}
\affil{DTU Space, National Space Institute, Technical University of Denmark, Elecktrovej 328, DK-2800 Kgs. Lyngby, Denmark}

\author[0000-0001-5727-4094]{Drake L. Deming}
\affil{Department of Astronomy, University of Maryland at College Park, College Park, MD 20742, USA} 

\author[0000-0002-1600-7835]{Jorge Sanz-Forcada}
\affil{Centro de Astrobiolog\'ia (CSIC-INTA), ESAC Campus, Villanueva de la Ca\~nada, Madrid, Spain}

\author[0000-0002-8507-1304]{Nikole K. Lewis}
\affil{Department of Astronomy and Carl Sagan Institute, Cornell University, Ithaca, NY 14853, USA}

\author[0000-0003-3726-5419]{Joanna K. Barstow}
\affil{School of Physical Sciences, The Open University, Walton Hall, Milton Keynes, MK7 6AA, UK}

\author[0000-0001-5442-1300]{Thomas Mikal-Evans}
\affil{Kavli Institute for Astrophysics and Space Research, Massachusetts Institute of Technology, 77 Massachusetts Avenue, 37-241, Cambridge, MA 02139, USA}

\author[0000-0003-1605-5666]{Lars A. Buchhave}
\affil{DTU Space, National Space Institute, Technical University of Denmark, Elecktrovej 328, DK-2800 Kgs. Lyngby, Denmark}

\begin{abstract}

\noindent Exoplanets with cloud-free, haze-free atmospheres at the pressures probed by transmission spectroscopy represent a valuable opportunity for detailed atmospheric characterization and precise chemical abundance constraints. We present the first optical to infrared (0.3$-$5 $\mu$m) transmission spectrum of the hot Jupiter WASP-62b, measured with \textit{Hubble}/STIS and \textit{Spitzer}/IRAC. The spectrum is characterized by a 5.1-$\sigma$ detection of Na {\sc i} absorption at $0.59\,\micron$, in which the pressure-broadened wings of the Na D-lines are observed from space for the first time. A spectral feature at $0.4\,\micron$ is tentatively attributed to SiH at 2.1-$\sigma$ confidence. Our retrieval analyses are consistent with a cloud-free atmosphere without significant contamination from stellar heterogeneities. We simulate \textit{James Webb Space Telescope} (JWST) observations, for a combination of instrument modes, to assess the atmospheric characterization potential of WASP-62b. We demonstrate that JWST can conclusively detect Na, H$_{2}$O, FeH, and SiH within the scope of its Early Release Science (ERS) program. As the only transiting giant planet currently known in the JWST Continuous Viewing Zone, WASP-62b could prove a benchmark giant exoplanet for detailed atmospheric characterization in the \textit{James Webb} era.

\end{abstract}

\keywords{planets and satellites: atmospheres  --- 
planets and satellites: composition --- planets and satellites: individual (WASP-62b)}

\section{Introduction} \label{sec:intro}


Close-in gas giant exoplanets that are cloud-free/haze-free in the observable atmosphere appear to be rare ($<$7\% of cases), with the vast majority of hot Jupiters showing substantial opacity from condensation clouds and photochemical hazes \citep{Wakeford19}. The few benchmark cases of clear atmospheres, such as WASP-96b \citep{Nikolov18} and WASP-39b (\citealt{Nikolov16,Fischer16,Wakeford18,Kirk19}), have produced some of the best constraints on atmospheric metallicities and abundances of H$_{2}$O and Na to date (e.g., \citealt{Welbanks19}). The rare class of cloud-free planets at the pressures probed via low-resolution transmission spectroscopy, ($\sim$10$^{-3}-$1 bar for optical and infrared observations) permit precision measurements of atomic and molecular abundances, unhindered by cloud-composition degeneracies (e.g., \citealt{Fraine14,Kreidberg14,Sing16,Kilpatrick18}). Exoplanets with clear atmospheres therefore represent a valuable opportunity to unlock crucial insights into atmospheric chemistry and planetary formation history (e.g., \citealt{Oberg11,Mordasini16}).  


The upcoming \textit{James Webb Space Telescope} (JWST) will enable unprecedented detailed atmospheric characterization of exoplanets, exploring a wider wavelength range ($0.6-28.3\,\micron$) than currently accessible with existing facilities \citep{Beichman14}. Its broad infrared wavelength coverage will allow precise molecular abundance constraints for many species, breaking degeneracies between parameters such as metallicity and the carbon-to-oxygen ratio. In preparation for the launch of JWST, the exoplanet community has invested significant effort in identifying planets that are cloud-free/haze-free in the observable atmosphere for follow-up studies with JWST \citep[see e.g.,][]{Stevenson16}. One of the targets initially put forth is WASP-62b \citep{Hellier12}, a 0.57 $M_{Jup}$, 1.39 $R_{Jup}$ planet with $T_{\rm eq} \sim 1440$\,K orbiting a bright ($V$ = 10.2) F7V host star. Due to its fortuitous location in JWST's Continuous Viewing Zone (CVZ), i.e. near the south ecliptic pole, WASP-62b was identified as a potential target for the Transiting Exoplanet Community Early Release Science Program (ERS 1366; P.I.: N. Batalha) for JWST \citep{Bean18}. 

Although WASP-62b is one of the most favorable planets for JWST atmospheric studies, another target (WASP-79b, which is not located in JWST's CVZ) was chosen for the ERS program due to a lack of observational information about the atmospheric properties of WASP-62b. To date, WASP-62b is the only known transiting giant planet in the CVZ. Although \textit{TESS} has found five giant planet candidates\footnote{Of the five \textit{TESS} CVZ candidates, four are fainter than WASP-62 (with $V$-band magnitudes ranging from 11.3 to 12.4). The candidate brighter than WASP-62 at $V$ = 8.6 mag has a V-shaped light curve and shows evidence of blending.} near the south ecliptic pole at the time of writing, none have been confirmed as planets (S. Quinn, priv. comm.). As a CVZ target, WASP-62b provides the opportunity to ensure that a suitable target is observable, regardless of any past or future JWST launch delays, with transit observations that can be flexibly scheduled and quickly executed.




In this Letter, we present the first optical transmission spectrum of the hot Jupiter WASP-62b. Our \textit{Hubble}/STIS and \textit{Spitzer}/IRAC observations demonstrate that this planet possesses a cloud-free terminator, rendering it a priority target for the JWST ERS Program. In what follows, we describe the data reduction and atmospheric retrieval of our new observations. We then discuss the implications of our atmospheric inferences, and provide testable predictions for infrared observations with JWST.


\section{Observations and Data Reduction}
\label{sec:obs}
We observed three transits of WASP-62b with STIS (\S \ref{sec:stis}) as part of the \textit{Hubble} Panchromatic Comparative Exoplanetology Treasury (PanCET) program GO~14767 (PIs: Sing \& L\'opez-Morales). We observed two additional transits with \textit{Spitzer}/IRAC (\S \ref{sec:irac}) in the $3.6\,\micron$ and $4.5\,\micron$ channels through GO~13044 (PI: Deming). Although \textit{Hubble}/WFC3 also observed one transit of this target for program GO~14767, we note that these observations suffered from guide star issues that render the data unreliable (\S \ref{sec:wfc3}).

\subsection{STIS}
\label{sec:stis}
The \textit{Hubble}/STIS transit observations consist of low-resolution ($\sim$500) time series spectra collected on UT 2018 Jan 26 (visit 57) and UT 2018 May 12 (visit 58) with the G430L (2892-5700 \AA) grism, and on UT 2017 Nov 21 (visit 59) with the G750L (5240-10270 \AA) grism. Each visit consisted of five consecutive 96-minute orbits, during which 48 stellar spectra were obtained over exposure times of 253 seconds. To decrease the readout times between exposures, we used a 128 pixel wide sub-array. The data were taken with the 52 x 2 arcsec$^{2}$ slit to minimize slit light losses. We note that both of the G430L visits suffered from guide star issues at the beginning of the observations, resulting in the loss of data collection during the first orbit (visit 57) or part of the first orbit (visit 58). From the subsequent exposures taken during these visits, however, we were able to extract good quality light curves with a median precision of 311 ppm.  

We reduced the STIS spectra using the methods described in \citet{Alam18,Alam20}. Briefly, we bias-, dark-, and flat-field corrected the raw 2D data frames using the CALSTIS pipeline (V 3.4). We corrected for cosmic ray events using median-combined difference images to flag and interpolate over bad pixels. We performed a 1D spectral extraction from the calibrated {\tt .flt} files and extracted light curves using an aperture width of 13 pixels. From the {\tt x1d} files, we obtained a wavelength solution by re-sampling all of the extracted spectra and cross-correlating them to a common rest frame. The cross-correlation measures the shift of each stellar spectrum with respect to the first spectrum of the time series, so we re-sampled the spectra to align them and remove sub-pixel drifts associated with the different locations of the spacecraft on its orbit \citep{Huitson13}.

\begin{figure*}
    \centering
    \includegraphics[width=0.9\textwidth, trim={0.0cm 1.0cm 0.0cm 0.1cm}]{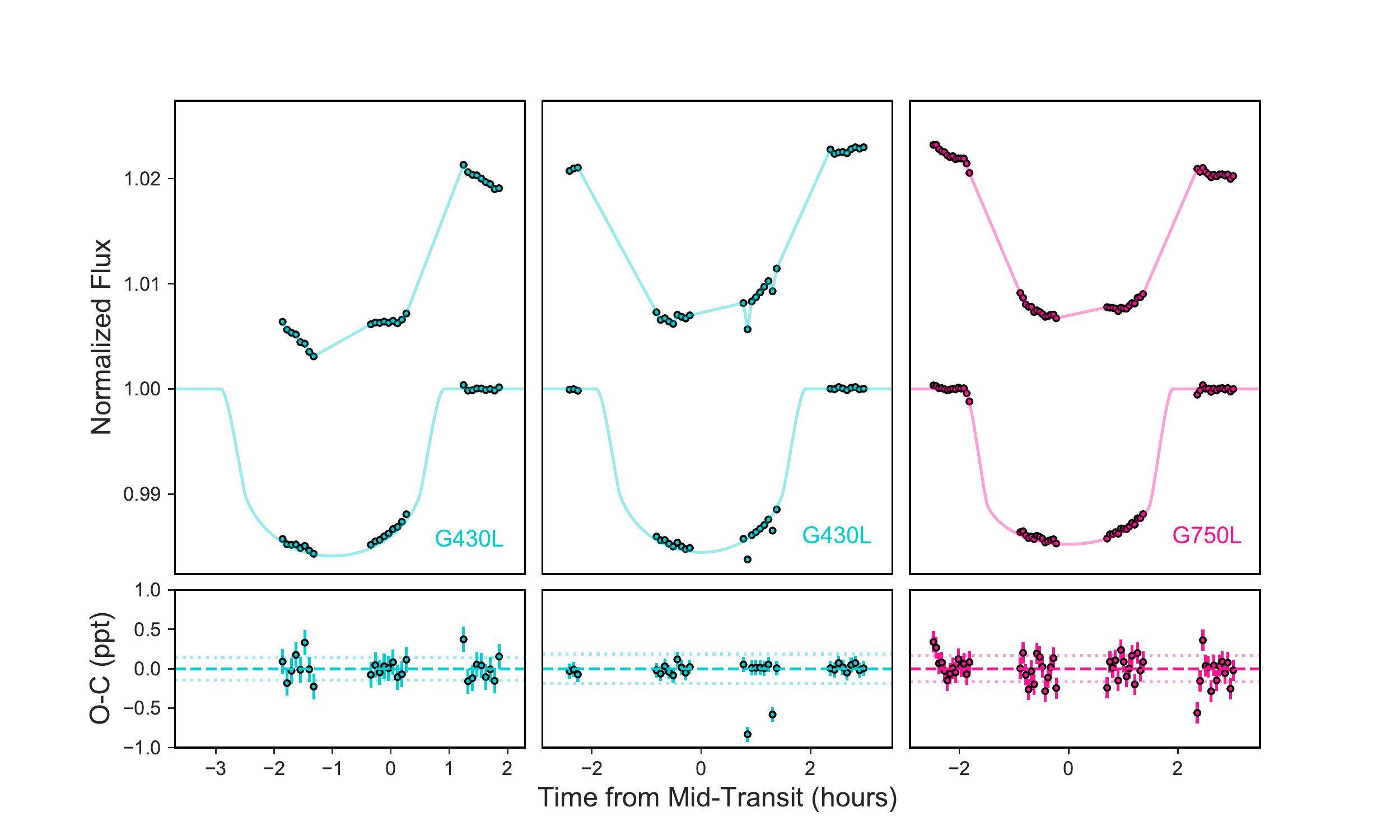}
    \caption{\textit{Top:} The raw and detrended white light curves (excluding the first orbit and the first exposure of each subsequent orbit) for each \textit{HST} visit in the STIS G430L (blue) and G750L (pink) gratings. The best-fit analytical light curve model is overplotted. \textit{Bottom:} Transit fit residuals (in parts per thousand) with error bars.}
    \label{fig:HST_wlc}
\end{figure*}

\subsection{IRAC}
\label{sec:irac}
We observed two transits with \textit{Spitzer}/IRAC on UT 2016 Nov 24 and UT 2016 Dec 7 in the $3.6\,\micron$ and $4.5\,\micron$ channels, respectively (\citealt{Werner04,Fazio04}). Each IRAC exposure was taken over integration times of 2 seconds, resulting in 20,159 images. We reduced the \textit{Spitzer} photometry using the techniques of \citet{Nikolov15}, \citet{Sing15,Sing16}, and \citet{Alam18}. We filtered for outliers in the data and subtracted the sky background using an iterative 3-$\sigma$ clipping procedure, as outlined in \citet{Knutson12} and \citet{Todorov13}. 

We extracted photometric points using two approaches: fixed and time-variable aperture photometry. In the fixed approach, we used circular apertures with radii ranging from 4 to 8 pixels in increments of 0.5. For the time-variable approach, we scaled the size of the aperture by the noise pixel parameter (the normalized effective background area of the IRAC point response function) which depends on the FWHM of the stellar PSF squared (\citealt{Mighell05,Knutson12,Lewis13,Nikolov15}). We compared the light curve residuals and the white and red noise components measured with the \citet{Carter09} wavelet technique to identify the best results from both methods.


\subsection{Unusable WFC3 Observations}
\label{sec:wfc3}

We also observed one transit of WASP-62b with \textit{Hubble}/WFC3 on UT 2017 April 14 (GO: 14767). This visit, however, suffered from severe guide star issues that impacted the data quality of the observations, rendering it unsuitable for reliably extracting information on the planet's atmosphere. Guide star issues during the planet's transit impacted the telescope guiding, resulting in large drifts on the order of several pixels that affect the photometric quality of the spectroscopic channels and contribute to systematic trends that drift in wavelength during the observation (see Figure 11 of \citealt{Sing18}). Further, the G141 stellar spectra exhibit large offsets in the stellar spectral structure, even though the edges of the spectrum are sharp, likely due to warping of the scan as the scan rate changes across the detector as well as positional shifts in the scan. Although a spectrum extracted from this observation was presented in \citet{Skaf20}, we do not use this G141 dataset in our analysis for the reasons outlined above. 

\subsection{Host Star X-Ray \& UV Monitoring}
\label{sec:photometry}

We were awarded XMM-Newton time (program ID 80479, P.I. J. Sanz-Forcada) to observe WASP-62 on UT 2017 April 14 for $\sim$8~ks. We fit European Photon Imaging Camera (EPIC) spectra using a one-temperature coronal model to calculate an X-ray luminosity (0.12$-$2.48 keV) of $L_{\rm X}=6.0\times 10^{28}$~erg\,s$^{-1}$ (S/N=5.8), for a Gaia DR2 distance of 176.53$\pm$0.41 pc. The resulting $\log L_{\rm X}/L_{\rm bol}=-5.1$ measurement indicates that the star has a moderate activity level, similar to the young solar analog $\iota$~Hor (G0V) in which X-ray variability is present but flares are not frequently observed \citep{Sanz19}. 

Furthermore, the X-ray and UV light curves taken with the XMM-Newton EPIC ($\sim$ 1$-$124~\AA) and the Optical Monitor (OM)/UVM2 filter (2070$-$2550~\AA) show some level of variability\footnote{The stellar X-ray and UV monitoring data are included in the online supplementary information.}. The UV light curve shows some dips (up to $\sim10\%$ absorption) around the beginning of the primary transit that could indicate the presence of occulted chromospheric plages as the planet transits. (Further details will be included in Sanz-Forcada et al., in prep.). Considering the activity level of the host star, we therefore account for starspots and faculae in our retrieval analysis (\S \ref{sec:retrievals}).


\section{Hubble and Spitzer Light Curve Fits}
\label{sec:analysis}
\subsection{Hubble}

We fit the light curves following the procedure detailed in \citealt{Kirk2017,Kirk2018,Kirk19}, which we briefly describe here. We modeled the analytic transit light curves of \cite{MandelAgol} using the \texttt{Batman} package \citep{batman}, combined with a Gaussian process (GP) implemented with the \texttt{george} \citep{george} code to model noise in the data. GPs are increasingly used in transmission spectroscopy \citep[e.g.,][]{Gibson2012,Gibson2012b,Evans2015,Evans2017,Kirk2017,Kirk2018,Kirk19,Louden2017,Parviainen2018}. 

For both the white light and spectroscopic light curve fits, we used the four parameter non-linear limb darkening law \citep{Claret00}. We derived the limb darkening coefficients using 3D stellar models \citep{Magic15}, and fixed them to the theoretical values in our fits. For the white light curve fits, we fixed the inclination $i$ to 88.3$^{\circ}$, the scaled semi-major axis $a/R_\star$ to 9.53, and the period $P$ to 4.4 days (based on the literature values from \citealt{Hellier12}), and fit for the time of mid-transit $T_0$, planet-to-star radius ratio $R_p/R_\star$, and the GP hyperparameters. Our GP was defined by the combination of 11 squared exponential kernels, each operating on a jitter detrending vector\footnote{As described in \citet{Sing19}, the jitter vectors found to highly correlate with \textit{Hubble}/STIS data are the right ascension and declination of the aperture reference as well as the roll of the telescope along the V2 and V3 axes.}. These 11 kernels each had their own length scale but shared a common amplitude. We additionally included a white noise kernel, defined by the variance ($\sigma^2$), to account for white noise unaccounted by the photometric error bars. The GP therefore added an additional 13 free parameters for each light curve. 

Following similar studies \citep[e.g.,][]{Evans2017,Evans2018}, we standardized each GP input variable (jitter detrending variable) by subtracting the mean and dividing by the standard deviation. This method gives each standardized variable a mean of zero and a standard deviation of unity, which helps the GP to determine the inputs of importance for describing the noise characteristics. We fit for the natural log of the inverse length scale \citep[e.g.,][]{Evans2017,Evans2018,Gibson2017}. We placed wide, truncated uniform priors in log space on each of the hyperparameters. This choice assures uninformative priors since a uniform prior in log-space is akin to fitting for a 1/x prior in non-logarithmic space. Since we are fitting for the natural log of the inverse length scale, this choice encourages the length scale to longer length scales so as to ensure that the GP does not over-fit the data (e.g., \citealt{Parviainen2018,Evans2018,Gibson19}). The GP amplitude was bounded between 0.01 and 100$\times$ the variance of the out-of-transit data, and the length scales were bounded by the typical spacing between data points and 5$\times$ the maximum length scale. The white noise variance was bounded between 10$^{-10}$ and 25 ppm. 


\begin{deluxetable}{cc}
\tabletypesize{\normalsize}
\tablewidth{0pt}
\tablecolumns{2}
\tablecaption{Transmission spectrum of WASP-62b measured with STIS G430L \& G750L and \textit{Spitzer} IRAC  
\label{tab:trspec}}

\tablehead{\colhead{$\lambda$ (\AA)}  & \colhead{$R_{p}/R_{*}$}  } 

\startdata 
2900$-$3700  & 0.11210 $_{-0.00115}^{+0.00116}$ \\ 
3700$-$4100  & 0.11285 $_{-0.00066}^{+0.00072}$ \\ 
4100$-$4400  & 0.11308 $_{-0.00051}^{+0.00053}$ \\ 
4400$-$4600  & 0.11219 $_{-0.00053}^{+0.00054}$ \\ 
4600$-$4700  & 0.11126 $_{-0.00053}^{+0.00050}$ \\ 
4700$-$4800  & 0.11212 $_{-0.00044}^{+0.00043}$ \\ 
4800$-$4900  & 0.11159 $_{-0.00049}^{+0.00053}$ \\ 
4900$-$5000  & 0.11133 $_{-0.00067}^{+0.00070}$ \\ 
5000$-$5100  & 0.11166 $_{-0.00059}^{+0.00055}$ \\ 
5100$-$5300  & 0.11199 $_{-0.00039}^{+0.00042}$ \\ 
5300$-$5500  & 0.11263 $_{-0.00063}^{+0.00052}$ \\ 
5500$-$5700  & 0.11184 $_{-0.00049}^{+0.00049}$ \\ 
5700$-$5800  & 0.11397 $_{-0.00078}^{+0.00069}$ \\ 
5800$-$5878  & 0.11362 $_{-0.00068}^{+0.00066}$ \\ 
5878$-$5913  & 0.11494 $_{-0.00066}^{+0.00063}$ \\ 
5913$-$6070  & 0.11290 $_{-0.00065}^{+0.00065}$ \\ 
6070$-$6200  & 0.11227 $_{-0.00051}^{+0.00047}$ \\ 
6200$-$6300  & 0.11299 $_{-0.00086}^{+0.00079}$ \\ 
6300$-$6513  & 0.11376 $_{-0.00059}^{+0.00056}$ \\ 
6513$-$6613  & 0.11489 $_{-0.00078}^{+0.00076}$ \\ 
6613$-$6800  & 0.11208 $_{-0.00075}^{+0.00074}$ \\ 
6800$-$7000  & 0.11344 $_{-0.00058}^{+0.00060}$ \\ 
7000$-$7200  & 0.11130 $_{-0.00061}^{+0.00058}$ \\ 
7200$-$7590  & 0.11077 $_{-0.00042}^{+0.00041}$ \\ 
7590$-$7740  & 0.10910 $_{-0.00072}^{+0.00074}$ \\ 
7740$-$8100  & 0.11136 $_{-0.00082}^{+0.00089}$ \\ 
8100$-$8500  & 0.11346 $_{-0.00125}^{+0.00112}$ \\ 
8500$-$8985  & 0.11417 $_{-0.00102}^{+0.00107}$ \\ 
8985$-$10300 & 0.11530 $_{-0.00154}^{+0.00152}$ \\ 
360000 & 0.10697 $\pm$ 0.00152 \\ 
450000 & 0.10962 $\pm$ 0.00040 \\[2pt] 
\enddata
\end{deluxetable}

We ran a Markov chain Monte Carlo (MCMC) using the \texttt{emcee} Python package \citep{emcee} to perform the fitting. For all of the light curve fits, we began by optimizing the GP hyperparameters to the out-of-transit data to find the starting locations for the GP hyperparameters. The starting value for $R_p/R_\star$ was taken from \citet{Hellier12} and from visual inspection of the light curve for $T_0$. The chains were then initialized with a small scatter around these starting values. We ran the MCMC for 2000 steps with 300 walkers ($20 \times n_p$, where $n_p$ is the number of parameters) and calculated the 16th, 50th and 84th percentiles for each parameter after discarding the first 1000 steps as burn in. Following the \texttt{george} documentation\footnote{https://george.readthedocs.io/en/latest/}, we then ran a second chain with the walkers initiated with a small scatter around the 50th percentile values for another 2000 steps with 300 walkers and again discarded the first 1000 steps. We measure $R_{p}/R_{\star}$ values of $0.11176_{-0.001394 }^{+0.001374}$, $0.111961_{-0.000729}^{+0.000804}$, and $0.113189_{-0.000692}^{+0.000750}$, for visits 57, 58, and 59, respectively, from the white light curves (Figure \ref{fig:HST_wlc}).

To derive the spectroscopic light curves, we binned the G430L and G750L spectra into 12 and 17 spectrophotometric channels, which are listed in the first column of Table \ref{tab:trspec}. We then fit and detrended each spectrophotometric light curve following the same procedure as the white light curves, but kept $T{_0}$ fixed to the result from the white light curve fit. We ran MCMCs to each light curve, following the same process as for the white light curves but with 280 walkers since there was one fewer fit parameter. The resulting ${R_{p}/R_{\star}}$ values  for  each  spectroscopic channel are presented in Table \ref{tab:trspec}\footnote{Figures of the white light curves and spectroscopic light curves are available online in the supplementary material.}. 
We used the values of $R_{p}/R_{\star}$ given in Table \ref{tab:trspec} to derive the optical transmission spectrum shown in Figure \ref{fig:STIS_Spitzer_retrieval}.

\subsection{Spitzer}

 We fit the 3.6 $\mu$m and 4.5 $\mu$m IRAC light curves following the methods described in \citet{Alam18}. Briefly, we corrected for flux variations from intra-pixel sensitivity (e.g., \citealt{Charbonneau05,Charbonneau08,Reach05,Knutson08,Ingalls12,Krick16}), and accounted for systematics by fitting a model of the functional form: 
\begin{equation}
F(t) = c_{0}~ + c_{1}x~ + c_{2}x^{2}~ + c_{3}y~ + c_{4}y^{2}~ + c_{5}xy~ + c_{6}t
\label{eq:irac}
\end{equation}
where $F(t)$ is the stellar flux as a function of time, the coefficients $c_{0}$ through $c_{6}$ are free parameters, $x$ and $y$ are the stellar centroid positions on the detector, and $t$ is time. We marginalized over all possible combinations of this model using the \citet{Gibson14} procedure. 
To measure $R_{p}/R_{\star}$ for the transmission spectrum, we fixed $P$, $a/R_{\star}$, and $i$ to the values from \citet{Hellier12} and fit for $R_{p}/R_{\star}$ and \textit{T}$_{0}$. The limb darkening coefficients were also fixed to their theoretical values based on 3D stellar atmosphere models. The measured $R_{p}/R_{\star}$ values are included in Table \ref{tab:trspec}. 
We note that our transit depth measurements are slightly lower than those previously published in \citet{Garhart20}, and this discrepancy may be due to differences in the data reduction procedure, such as choice of baseline ramp shapes (linear versus quadratic) or where to trim out-of-transit data.



\section{Atmospheric Retrievals}
\label{sec:retrievals}

\begin{figure*}
    \centering
    \includegraphics[width=0.9\textwidth, trim={0.0cm 1.0cm 0.0cm 0.1cm}]{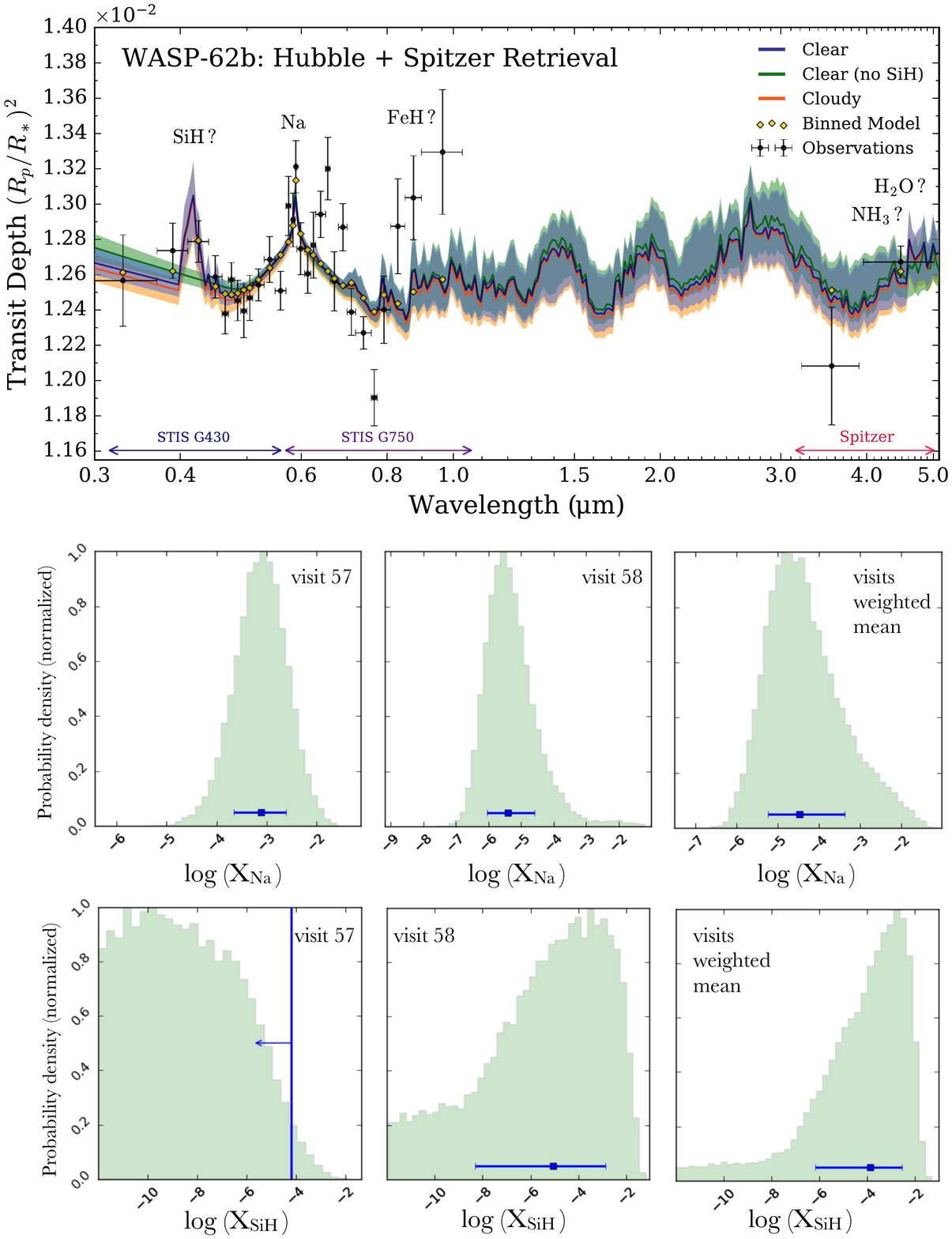}
    \caption{\textit{Top}: Transmission spectrum for WASP-62b measured with \textit{Hubble} and \textit{Spitzer} (black points), along with the best-fit model from our retrieval analysis (navy line) binned to the resolution of our observations (gold diamonds). The shaded regions indicate the 1-$\sigma$ confidence intervals for the best-fit retrieved spectrum (blue), a clear model excluding SiH opacity (green), and a model incorporating clouds (orange). \textit{Middle}: Posterior histograms comparing the retrieved Na abundance when including only STIS G430L visit 57 (left), only G430L visit 58 (center), and the weighted mean of the two G430L datasets (right). \textit{Bottom}: Same as above, but for the retrieved SiH abundances.}
    \label{fig:STIS_Spitzer_retrieval}
\end{figure*}


We retrieved the atmospheric properties of WASP-62b using the POSEIDON radiative transfer and retrieval code \citep{MacDonald17}. POSEIDON generates $\gtrsim 10^5 - 10^6$ model atmospheres -- spanning a wide range of chemical abundances, temperature structures, and cloud properties -- to identify the underlying atmospheric properties required to explain an observed exoplanet spectrum. Our initial exploration of WASP-62b's STIS+\textit{Spitzer} transmission spectrum considered 19 chemical species (Na, K, Li, H$^-$, TiO, VO, AlO, SiO, TiH, CrH, FeH, CaH, MgH, NaH, SiH, H$_2$O, CH$_4$, HCN, NH$_3$) known to present strong opacity at optical wavelengths \citep{Tennyson18}, collision-induced absorption due to H$_2$-H$_2$ and H$_2$-He pairs \citep{Karman19}, an isothermal temperature structure, and a patchy cloud/haze model \citep{MacDonald17}. Given the moderate stellar activity level indicated by the X-ray and UV photometric monitoring (\S \ref{sec:photometry}), we also include a prescription for contamination due to unocculted starspots or faculae \citep{Rackham18,Pinhas18}. We explore this 28-dimensional parameter space with the Bayesian nested sampling algorithm MultiNest (\citealt{Feroz08,Feroz09,Feroz19}), implemented by PyMultiNest \citep{Buchner14}, with 4,000 live points.

The best-fitting model from our atmospheric retrieval analysis is consistent with a clear atmosphere free from stellar contamination (Figure~\ref{fig:STIS_Spitzer_retrieval}). Na {\sc i} absorption, detected at 5.1-$\sigma$ confidence, characterizes WASP-62b's optical transmission spectrum. The observed pressure-broadened wings of the Na D-lines at $0.59\,\micron$ strongly favor a clear atmosphere, permitting a precise retrieved Na abundance: $\log (\rm{X_{Na}}) = -4.46_{-0.76}^{+1.09}$. The only other molecules with Bayes factors $>$ 1 are H$_2$O, NH$_3$, FeH, and SiH, with at least one of these species required at 2.8-$\sigma$ confidence. We tentatively attribute a spectral feature around $0.4\,\micron$ to SiH at 2.1-$\sigma$ confidence. 

The Bayesian evidences from models including clouds or stellar contamination were lower than the corresponding clear atmosphere models, indicating WASP-62b's observed transmission spectrum does not favor these features. We obtain a 2-$\sigma$ lower limit on the cloud top pressure of log($P_{cloud}$) $>$ -5.56. Based on this, we conclude that the observable atmosphere of WASP-62b is cloud-free at the pressures probed by our transmission spectra observations. We established, via successive Bayesian model comparisons, that the simplest model consistent with the present data is a clear, isothermal atmosphere with Na, H$_2$O, NH$_3$, FeH, and SiH. This 7-parameter model is compared to our STIS+\textit{Spitzer} transmission spectrum observations in Figure~\ref{fig:STIS_Spitzer_retrieval}.

The large scatter at the red end of the spectrum is responsible for the non-significant inference of FeH. We tested how these spectrophotometric channels affect the retrieval results by repeating our analysis with the last three STIS G750L wavelength bins excluded. The removal of these three reddest STIS points changes the FeH posterior from a bounded constraint into an upper limit, but all other results remain consistent. We further note that an L-shaped degeneracy in the H$_2$O-NH$_3$ correlation plot\footnote{Posterior distributions of retrieved parameters, and the transmission contribution function (\citealt{Molliere19}) for our best-fitting model, are available in the supplementary online material.} suggests that the data require at least one of these species, but their absorption signatures are degenerate with the data in hand. The presence of H$_2$O is indicated by the $4.5\,\micron$ \textit{Spitzer} channel, although further infrared observations are required to confirm H$_2$O opacity and precisely constrain its abundance.  

To investigate the reliability of our inferences, we also retrieved each STIS G430L visit separately (alongside the STIS G750L and \textit{Spitzer} data). We compare the retrieved Na and SiH abundances from each G430L visit to our baseline two-visit weighted mean in the lower panels of Figure~\ref{fig:STIS_Spitzer_retrieval}. Each visit separately detects the pressure-broadened Na wings, albeit with abundances discrepant by $\sim$ 2.5\,dex. In particular, the precise Na abundance from the visit 58 dataset, $\log (\rm{X_{Na}}) = -5.41_{-0.62}^{+0.82}$, is consistent with expectations for a solar metallicity atmosphere ($\log (\rm{X_{Na}})_{\Sun} = -5.76$, \citealt{Asplund09}). The tentative evidence of SiH seems to be driven by visit 58, which is more precise (mean error = 139\,ppm) than the visit 57 dataset (483\,ppm). However, the low abundance tail in the SiH posterior becomes less probable for the weighted mean of the two visits compared to visit 58 alone. This sharper SiH posterior for the combined visit retrieval suggests the spectra from each visit are not in tension, with the visit 57 retrievals not inferring SiH only due to the larger uncertainties. 

We note that our retrieved limb temperature for WASP-62b, $835_{-150}^{+188}$\,K, is markedly lower than both the equilibrium temperature ($1394_{-20}^{+25}$\,K) and skin temperature ($\sim 1170$\,K) of the planet. This low retrieved temperature is consistent with the recent results of \citet{MacDonald20}, who showed that applying 1D atmospheric models to transmission spectra with different morning-evening terminator compositions leads to cooler retrieved temperatures. We additionally validated the isothermal assumption for the present data by running a retrieval with the 6-parameter pressure-temperature profile of \citet{Madhusudhan09}. The results are consistent with those above, with only a marginal preference for a weak vertical temperature gradient at the terminator (Bayes factor = 1.4).

For a further test, we ran self-consistent retrievals using the ATMO Retrieval Code (ARC) under the assumptions of chemical equilibrium and local condensation (e.g., \citealt{Evans2017,Lewis20}). We ran one model including K and another model with K artificially removed (since the K feature is not observed in our data). Compared to the POSEIDON free retrieval results presented above, the self-consistent model finds somewhat hotter retrieved temperatures (1069$^{+85}_{-48}$ K without potassium; 1059$^{+100}_{-44}$ K with potassium) that are closer to the planet’s equilibrium temperature (1400 K). These are, however, consistent with the retrieved temperature from the free retrieval to within 1-$\sigma$.

The fit quality of the self-consistent model is strongly dependent on whether K absorption is included. Without K we find $\chi^{2}$ = 79.74, while this degrades to $\chi^{2}$ =  99.57 when K is included. This discrepancy suggests that some other process (besides local condensation) may be removing K from the gas phase. We note that the minimum $\chi^{2}$ somewhat prefers our minimal (7-parameter) free retrieval ($\chi^{2}$ = 60.48) over the self-consistent retrievals.

\section{Discussion}
\label{sec:discussion}

\subsection{A Clear Atmosphere for WASP-62b}
\label{sec:wasp62b_clear_discussion}

Due to the nearly ubiquitous nature of condensation clouds and photochemical hazes in exoplanet atmospheres (e.g., \citealt{Wakeford19}), few benchmark planets with cloud-free, haze-free atmospheres at the pressures probed by transmission spectroscopy are currently known. The detection of the Na {\sc i} line wings at 0.59 $\mu$m in the atmosphere of WASP-62b marks the first space-based observation of the pressure-broadened wings of the Na D-lines, and suggests that WASP-62b possesses a clear terminator. Clear atmosphere exoplanets present an unmatched opportunity to obtain increasingly precise retrieved abundance constraints, since they are unhampered by cloud-composition degeneracies (e.g., Figure 10 of \citealt{MacDonald17}). 

\begin{figure}[ht!]
    \centering
    \includegraphics[width=\columnwidth, trim={1.0cm 0.3cm 2.0cm 0.5cm}]{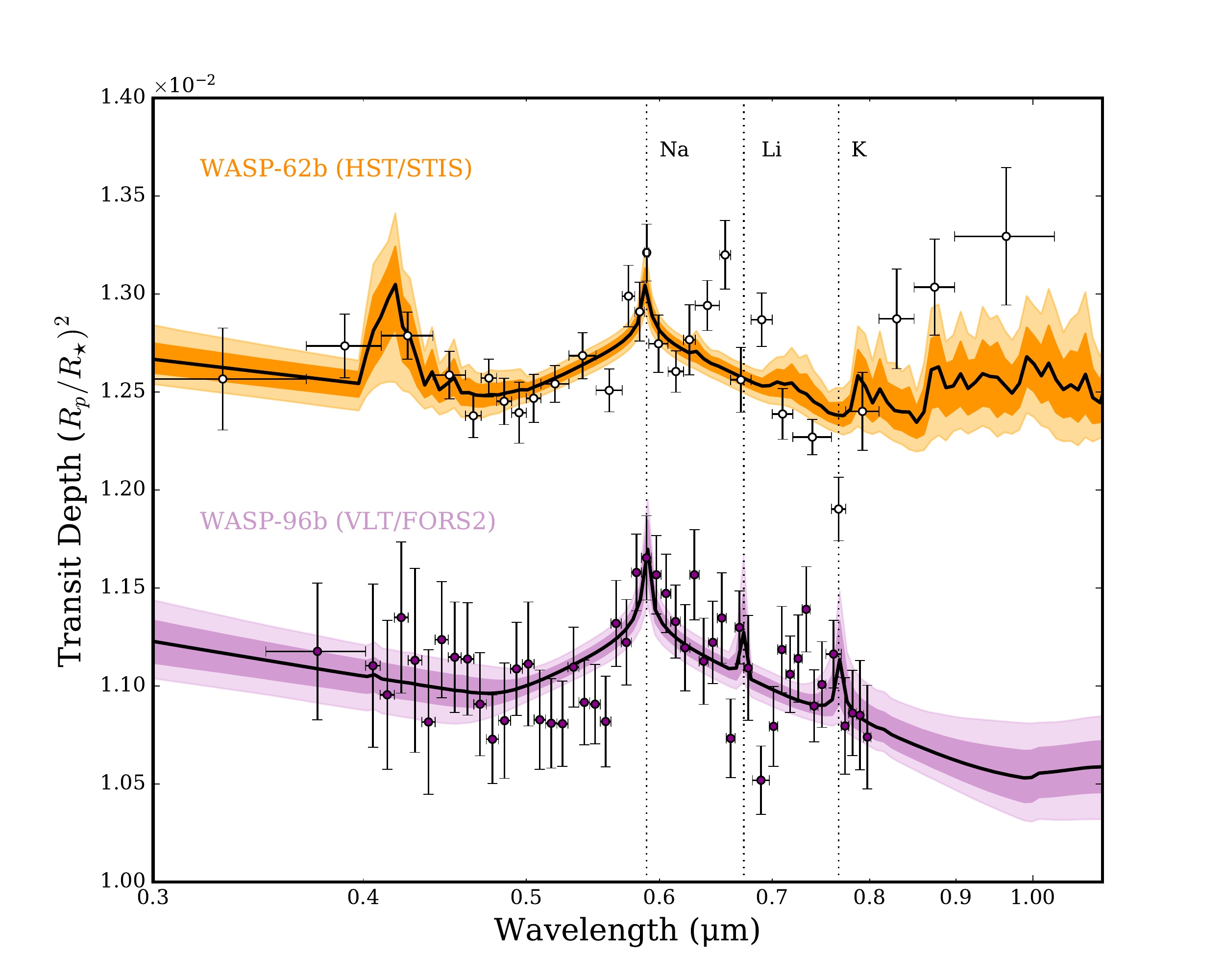}
    \caption{The \textit{Hubble}/STIS optical transmission spectra (white points) of WASP-62b in this work, compared with the VLT/FORS2 transmission spectrum (purple points) of WASP-96b (\citealt{Nikolov18}; offset vertically for clarity). The shaded regions show the 1-$\sigma$ and 2-$\sigma$ confidence intervals from cloud-free POSEIDON retrieval models for each planet. At low spectral resolution, both datasets display the prominent pressure-broadened wings of the Na D-lines at $0.59\,\micron$. WASP-96b additionally displays weak evidence of K {\sc i} at $0.77\,\micron$ and a hint of Li {\sc i} at $0.67\,\micron$. Our transmission spectrum of WASP-62b represents the first clear detection of pressure-broadened Na wings from a space-based telescope.}
    \label{fig:clear_planets}
\end{figure}

We compare our results for WASP-62b with the ground-based VLT/FORS2 observations of WASP-96b, the clearest known exoplanet to date \citep{Nikolov18}, in Figure \ref{fig:clear_planets}. Although both WASP-62b and WASP-96b display the pressure-broadened Na line wings, WASP-62b shows no evidence of K {\sc i} absorption at 0.76 $\mu$m whereas WASP-96b is consistent with weak evidence of K {\sc i} and Li {\sc i} \citep{Nikolov18}. The missing or weak potassium absorption for these clear planets is contrary to atmospheric models which predict the presence of both alkali features in clear exoplanets \citep{Seager00}. This observed trend may be due to a difference in the primordial abundances of the species, since Na is $\sim$15 times more abundant than K in a solar abundance atmosphere \citep{Lodders03}. The relative condensation temperatures of NaCl and KCl, as well as the photoionization energies of sodium and potassium, may further alter the relative amplitudes of the Na and K absorption features \citep{Nikolov18}. 

\subsection{ATMO Predictions for Si-bearing Species}
\label{sec:atmo_predictions}

In light of our tentative inference of SiH opacity in WASP-62b's transmission spectrum, here we consider theoretical predictions for gas-phase Si-bearing molecules at the equilibrium temperature ($T_{\rm eq} = 1440 \pm 30$\,K; \citealt{Hellier12}) of WASP-62b. Equilibrium chemistry expectations for silicates suggest that the dominant Si-bearing species are SiO and SiS at $\sim 1400$\,K \citep{Visscher10,Woitke18}. We investigate predictions for Si-bearing species in the atmosphere of WASP-62b using the planet-specific forward model ATMO grid (e.g., \citealt{Tremblin15,Goyal20}) of 1D radiative-convective equilibrium pressure-temperature profiles with corresponding self-consistent equilibrium chemistry. This grid of model atmospheres spans a range of recirculation factors (RCF), metallicities, and C/O ratios\footnote{We produced a range of C/O ratios by varying the oxygen abundance O/H, as detailed in \S 4 and \S 2.3 of the supplementary material of \citet{Goyal20}.}. The RCF parameterizes the redistribution of input stellar energy in the planetary atmosphere, where a value of $0.5$ represents efficient redistribution and $1$ corresponds to no redistribution.

\begin{figure}
    \centering
    \includegraphics[width=\columnwidth, trim={1.0cm 0.0cm 1.0cm -0.5cm}]{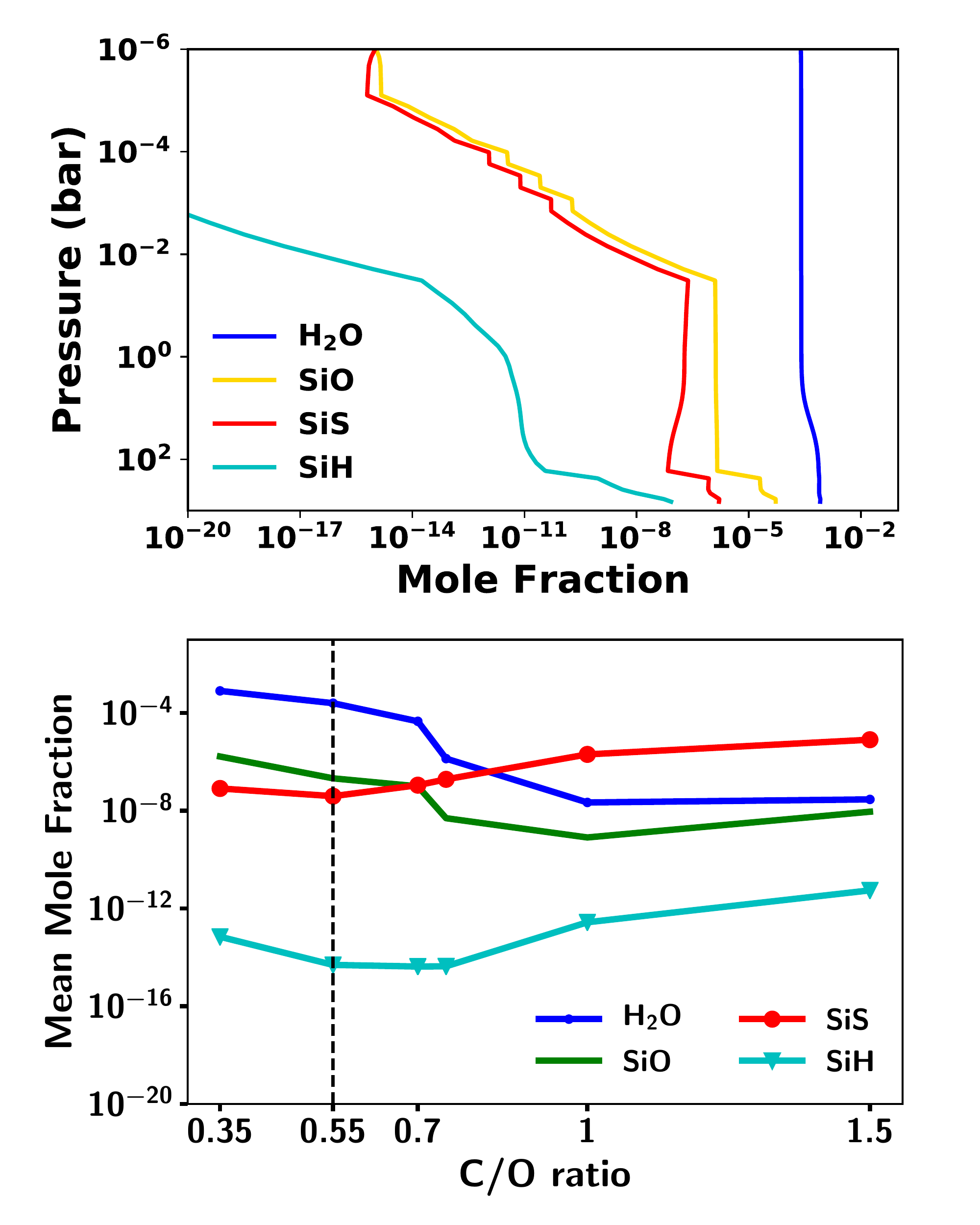}
    \caption{\textit{Top}: Theoretical abundances of H$_{2}$O (blue), SiO (yellow), SiS (red), and SiH (cyan) based on 1D equilibrium chemistry ATMO models of WASP-62b at solar metallicity. SiH is predicted to be less abundant than SiS or SiO at the equilibrium temperature of WASP-62b. \textit{Bottom}: Mean abundances (between 0.1 and 100 mbar) for these molecules as C/O varies. The vertical dashed line denotes a solar C/O. The abundances of SiO and H$_{2}$O drop by orders of magnitude for even moderately super-solar C/O, with SiS predicted to be the dominant Si-bearing molecule for C/O $\gtrsim 0.7$.}
    \label{fig:Si_abundances}
\end{figure}

Figure~\ref{fig:Si_abundances} (top panel) shows the equilibrium chemical abundances of H$_{2}$O, SiH, SiO, and SiS with uniform re-distribution (RCF = 0.5), solar metallicity, and a solar C/O ratio. While SiH is tentatively inferred by our retrievals, equilibrium expectations predict that SiO and SiS should be orders of magnitude more abundant than SiH in WASP-62b's atmosphere. Since silicate condensation has also been found to increase the carbon-to-oxygen (C/O) ratio in gas from solar to super-solar values \citep{Woitke18}, we also investigate how the abundances of these species change for sub-solar, solar, and super-solar C/O. Figure~\ref{fig:Si_abundances} (bottom panel) shows that SiS overtakes SiO to become the most abundant Si-bearing molecule for C/O $\gtrsim 0.7$. This is driven by a decrease in the SiO abundance for super-solar C/O ratios, analogous to the well-known decrease in H$_{2}$O abundance for enhanced C/O ratios \citep{Madhusudhan12}. These results expand upon the silicon chemistry predictions from \citet{Visscher10} - who considered rainout Si chemistry at a solar C/O ratio - demonstrating that SiS may be the most abundant gas-phase Si-bearing species for exoplanets with super-solar C/O ratios.


These equilibrium predictions appear in tension with our inferred SiH abundance ($\sim 10^{-4}$). Our retrievals do not infer SiO, despite its inclusion, due to its characteristic slope at UV-optical wavelengths \citep[e.g.][]{Sharp07} differing from the feature at $0.4\,\micron$ we attribute to SiH. Evidence of SiS cannot currently be assessed at optical wavelengths, due to the lack of currently available line lists for low energy (bluer than $\sim 2\,\micron$) transitions \citep{Upadhyay18}. It is therefore possible that the spectral feature we attribute to SiH could be a misclassified SiS feature - optical line list data for SiS would resolve this ambiguity. However, we stress that the evidence for SiH from our STIS data remains low (2.1-$\sigma$), and future observations will be required to assess the presence of gas-phase Si-bearing species in WASP-62b's atmosphere.



\subsection{Predictions for JWST}
\label{sec:sim_retrievals}

\begin{figure*}
    \centering
    \includegraphics[width=0.975\textwidth]{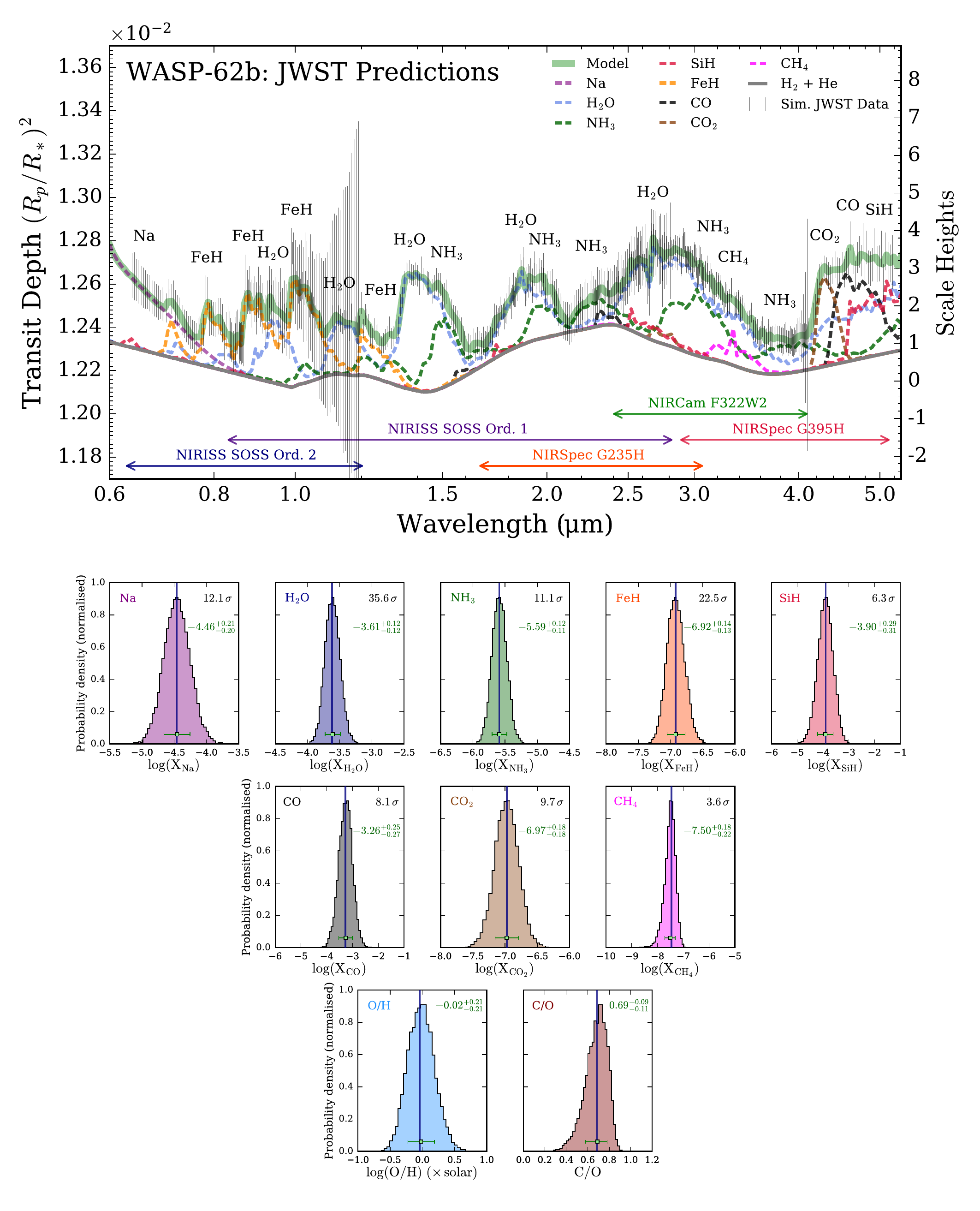}
    \vspace{-0.4cm}
    \caption{\textit{Top}: Simulated JWST observations for NIRSpec G235H+G395H, NIRCam F322W2, and NIRISS SOSS orders 1 \& 2 modes (gray points), for a single transit of WASP-62b with each mode, along with the best-fitting retrieved model (solid green line) and associated opacity contributions from Na, H$_{2}$O, NH$_{3}$, SiH, FeH, CO, CO$_{2}$, CH$_{4}$ (dashed lines), and the H$_{2}$+He continuum (solid gray line). \textit{Bottom}: Retrieved abundance posteriors for Na, H$_{2}$O, NH$_{3}$, FeH, SiH, CO, CO$_{2}$, CH$_{4}$, metallicity (O/H) and C/O. The abundances used to generate the model (solid vertical lines) agree well with the retrieved abundances (green error bars). JWST can conclusively detect and obtain precise abundance constraints for all the optical and infrared absorbers included in our model within the scope of the JWST ERS program.} 
    \label{fig:simulated_JWST_ERS}
\end{figure*}

A clear atmosphere for WASP-62b opens the door to extreme precision molecular abundance measurements from early JWST observations. To quantitatively assess the potential of WASP-62b as a priority JWST target, we offer predictions from a simulated retrieval analysis of synthetic JWST ERS observations of WASP-62b.

We generated a moderate-resolution ($R = 10,000$) model transmission spectrum, including Na, H$_2$O, NH$_3$, FeH, and SiH, with the median retrieved atmospheric properties from our HST+\textit{Spitzer} analysis. To explore the ability of JWST to constrain WASP-62's C/O ratio, we additionally injected CO, CO$_{2}$, and CH$_{4}$ abundances into our model. The representative values were taken to be the 10 mbar abundances of these species from the self-consistent model grid of \citet{Goyal20}, assumed uniform throughout the atmosphere. We then simulated JWST ERS observations with PandExo\footnote{\url{https://natashabatalha.github.io/PandExo/}} \citep{Batalha17} -- in the Panchromatic Transmission configuration outlined by \citet{Bean18} -- spanning four transits across the following modes:

\begin{itemize}
    \item NIRISS SOSS orders 1 \& 2 ($0.64 - 2.8\,\micron$) [61 ppm / 141 ppm].
    \item NIRSpec G235H ($1.7 - 3.1\,\micron$) [39 ppm].
    \item NIRCam F322W2 ($2.4 - 4.1\,\micron$) [67 ppm].
    \item NIRSpec G395H ($2.9 - 5.2\,\micron$) [57 ppm].
\end{itemize}

We binned the simulated PandExo observations to $R = 100$ for all modes, neglecting errors exceeding 1,000\,ppm (mainly the reddest NIRISS SOSS 2nd order data and near the NIRSpec detector gaps at $2.24\,\micron$ and $3.80\,\micron$), for a total of 677 data points. The square brackets above denote the resulting mean data precision for each mode. We removed Gaussian scatter from the dataset, centering the data on the actual model, to ensure our retrieval results are unbiased by a given noise instance \citep[see][]{Feng18}. Finally, we retrieved the synthetic dataset using POSEIDON.

Our simulated JWST observations and best-fitting retrieved model spectrum are shown in Figure~\ref{fig:simulated_JWST_ERS}. The combined ERS instrument modes can detect many prominent optical and infrared spectral features, as well as obtain precise constraints on the planetary atmospheric metallicity (here taken as O/H) and C/O. Each JWST ERS mode covers at least one of the H$_2$O band heads at $0.95\,\micron$, $1.15\,\micron$, $1.4\,\micron$, $1.9\,\micron$, $2.7\,\micron$, and $4.3\,\micron$. NIRISS SOSS samples both the red wings of the Na D-lines and strong FeH features at $0.8\,\micron$, $0.9\,\micron$, and $1.0\,\micron$. Both NIRISS SOSS and NIRSpec G235H are highly sensitive to NH$_3$ absorption, via the K-band feature at $2.2\,\micron$ \citep[see also][]{MacDonald17b}. NIRCam F322W2 and NIRSpec G395H sample the CO, CO$_2$, and CH$_4$ features between $3-5\,\micron$. NIRSpec G395H can also probe a broad SiH feature from $\sim 4.6 - 5.3\,\micron$, testing if gas-phase silicon species are indeed present in WASP-62b's atmosphere. The combined observing configuration also provides the optical baseline necessary for obtaining precise molecular abundances and atmospheric metallicities.

We predict that JWST observations of WASP-62b, within the scope of the ERS program, can conclusively detect Na (12.1-$\sigma$), H$_2$O (35.6-$\sigma$), FeH (22.5-$\sigma$), SiH (6.3-$\sigma$), NH$_{3}$ (11.1-$\sigma$), CO (8.1-$\sigma$), CO$_{2}$ (9.7-$\sigma$), and CH$_{4}$ (3.6-$\sigma$). The clear atmosphere offers remarkably precise abundance constraints: 0.12\,dex for H$_2$O, 0.14\,dex for FeH, 0.21\,dex for Na, and 0.30\,dex for SiH, 0.26\,dex for CO, 0.18\,dex for CO$_{2}$, and 0.20\,dex for CH$_{4}$ (shown in Figure~\ref{fig:simulated_JWST_ERS}, lower panels). These predicted abundance constraints, from a single transit with each JWST mode, would immediately outclass the most precise abundances obtained for hot Jupiters with existing facilities to date ($\lesssim$ 0.3\,dex for H$_2$O, e.g. \citealt{Welbanks19}). 


One of the key goals of JWST is to measure the metallicity and C/O ratios of a population of exoplanets \citep{Beichman14,Bean18}.
Our injected abundances of CO, CO$_2$, and CH$_4$ allow us to explore the expected constraints achievable by the JWST ERS program. We derive posterior distributions for the metallicity (O/H in solar units) and C/O from our individual abundance posteriors using the methodology of \citet{MacDonald19}, as shown on the bottom row of Figure~\ref{fig:simulated_JWST_ERS}. We predict that it is possible to constrain WASP-62b's C/O to $\pm 0.1$ and the atmospheric metallicity to $\pm 0.21$\,dex. Note that these constraints do not require the assumption of chemical equilibrium. 

\section{Summary \& Conclusions}
\label{sec:summary}

We presented the STIS+\textit{Spitzer} transmission spectrum of WASP-62b, the only transiting giant planet currently known in the JWST CVZ. WASP-62b is one of the few exoplanets with observed pressure-broadened alkali line wings, and the first space-based transmission spectrum displaying pressure-broadening Na {\sc i} absorption. Our retrievals are consistent with a cloud-free, haze-free atmosphere, with a strong detection of Na {\sc i} at 5.1-$\sigma$ confidence and tentative evidence of SiH at $0.4\,\micron$. 

We explored the prospects for JWST observations of WASP-62b via a simulated retrieval exercise. Conclusive detections ($>$ 5-$\sigma$) of Na, H$_2$O, FeH, and SiH can be achieved within the scope of the JWST ERS program, with the clear nature of WASP-62b's atmosphere offering abundance constraints at $< 0.2$\,dex precision. If confirmed, gas-phase SiH features observed near $5\,\micron$ would complement observations of condensed silicates (e.g., \citealt{Gao20}) via vibrational mode resonance features - accessible with JWST's MIRI LRS mode at longer wavelengths \citep{Wakeford15}. These results suggest WASP-62b is an exceptional target for JWST transmission spectroscopy. 



In preparation for JWST, identifying targets that are cloud-free/haze-free is important for mobilizing community efforts to observe the best planets for detailed atmospheric follow-up. Although alternative targets have since been put forward, WASP-62 is the only star in the JWST CVZ with a known transiting giant planet that is bright enough for high-quality atmospheric characterization via transit spectroscopy. JWST transit programs require many repeated visits, which ideally could be scheduled at any time of the year and executed quickly. WASP-62b is therefore one of the most readily accessible targets for atmospheric studies with JWST.

\acknowledgments
The authors thank the referee for their insightful comments and suggestions. M.K.A. thanks Mark Marley for useful discussions. R.J.M. thanks Sergei Yurchenko for helpful discussions on molecular line lists. This paper makes use of observations from the NASA/ESA \textit{Hubble Space Telescope}, obtained at the Space Telescope Science Institute, which is operated by the Association of Universities for Research in Astronomy, Inc., under NASA contract NAS 5-26555. These observations are associated with the \textit{Hubble} GO program 14767. M.K.A. acknowledges support by the National Science Foundation through a Graduate Research Fellowship. J.S.F. acknowledges support from the Spanish State Research Agency project AYA2016-79425-C3-2-P. 

\software{Batman \citep{batman}, george \citep{george}, emcee \citep{emcee}, POSEIDON \citep{MacDonald17}}


\bibliography{sample63}
\bibliographystyle{aasjournal}

\end{document}